\documentclass[twocolumn,showpacs,preprintnumbers,amsmath,amssymb]{revtex4}

\usepackage{graphicx}
\usepackage{dcolumn}
\usepackage{bm}
\usepackage{color}

\begin{document}

\preprint{APS/123-QED}

\title{Spin vortices and vacancies: interactions and pinning on a square lattice}

\author{O. Kapikranian}
\email{akap@icmp.lviv.ua}
\affiliation{%
Institute for Condensed Matter Physics, National Academy of
Sciences of Ukraine, 79011 Lviv, Ukraine
}%

\author{Yu. Holovatch}
\affiliation{%
Institute for Condensed Matter Physics, National Academy of
Sciences of Ukraine, 79011 Lviv, Ukraine,\\ Institut f\"ur
Theoretische Physik, Johannes Kepler Universit\"at Linz, 4040
Linz, Austria
}%

\date{\today}

\begin{abstract}

The study gives a decisive answer to the recently risen question
about the type and origin of interaction between spin vortices and
spin vacancies in $2D$ spin models. The approach is based on the
low-temperature approximation of the $2D$ $XY$ model known as the
Villain model and does not involve any additional approximations,
thus preserving the lattice structure. The exact form of the
Hamiltonian describing a system of topological charges and a
vacant site supports the attractive type of interaction between
the vacancy and the charges.

The quantitative difference between the characteristics of the
vortex behavior in the $2D$ $XY$ and Villain models due to the
different energy of the vortex ``cores'' in the two models is
pointed out. This leads to a conclusion that the interaction
between a vortex and a spin vacancy and between a vortex and the
antivortex differs quantitatively for small separations in the two
mentioned models.
\end{abstract}

\pacs{05.50.+q; 75.10}
\maketitle

\section{\label{intro}Introduction}

The term \textit{spin vortex} has become common in theoretical and
experimental studies of magnetic materials. It is, in fact, a
particular case of a more general class of physical/mathematical
objects called \textit{topological
defects}~\cite{Mermin79,ChaikinLubenskiiBook}.

Although, strictly speaking, topological defects can be defined
only in terms of a continuous field, similar formations can be
observed in lattice spin models. Moreover, it is the spin vortices
that are responsible for the
\textit{Berezinskii-Kosterlitz-Thouless (BKT) phase transition} in
the $2D$ $XY$
model~\cite{Berezinskii72,KosterlitzThouless73,Kosterlitz74} (or,
speaking more generally, in classical $2D$ easy-plane magnets).

Most of the theoretical studies of the vortex properties are
limited to the low-temperature continuum model proposed by
Kosterlitz and Thouless \cite{KosterlitzThouless73} (KT model,
hereafter). However, this approach obviously cannot give
satisfactory results, when essentially ``discrete'' phenomena, as
the effects induced by a spinless site, are studied. The lack of
theoretical studies regarding spin vortices on a lattice and the
related problem of spin vortex--spin vacancy interaction is the
principal motivation for the present work.

\subsection{\label{intro-A} Spin vortices}

The $2D$ $XY$ model is usually defined as a system of
two-component spins ${\bf S}_{\bf r}$ of unit length which states
can be represented by a polar coordinate $-\pi < \theta\le \pi$:
${\bf S}_{\bf r} = (\cos\theta_{\bf r}, \sin\theta_{\bf r})$,
placed at sites $\bf r$ of a square lattice, and described by the
Hamiltonian
\begin{equation}\label{H_XY_0}
H_{2DXY} = J\sum_{\left<{\bf r, r'}\right>}[1-\cos(\theta_{\bf r}
- \theta_{\bf r'})]\ .
\end{equation}
Close enough to the ground state we have $\theta_{\bf r} -
\theta_{\bf r'} \simeq 0$ or $\pm 2\pi$ for neighboring sites $\bf
r$,~$\bf r'$.

Generally, considering two neighboring spins at sites $\bf r$ and
$\bf r'$, one can encounter the two situations: $|\theta_{\bf r} -
\theta_{\bf r'}| < \pi$ and $|\theta_{\bf r} - \theta_{\bf r'}|
> \pi$ (the situation $|\theta_{\bf r} -
\theta_{\bf r'}| = \pi$ can be neglected). In order to define spin
vortices in the system under consideration, let us introduce the
lattice of sites $\bf R$, dual to the original lattice (the dual
lattice is the set of all the centers of elementary cells of the
original lattice), and consider only those bonds $({\bf R,R'})$
which intersect the bonds $({\bf r,r'})$ of the original lattice
for which $|\theta_{\bf r} - \theta_{\bf r'}| > \pi$.

In order to consider the bonds of interest in a systematic way,
let us say that ${\bf r} = (x,y)$, ${\bf r'} = (x',y')$ define the
bond $({\bf r,r'})$ if ${\bf r'} = (x+a,y)$ for a horizontal bond
and ${\bf r'} = (x,y+a)$ for a vertical bond, where $a$ is the
lattice spacing. The same rule is imposed for bonds of the dual
lattice. Now, we can ascribe to every bond $({\bf R,R'})$ a
direction defined by the sign of $\theta_{\bf r} - \theta_{\bf
r'}$ of the intersected bond $({\bf r,r'})$: $({\bf R \to R'})$ if
$\theta_{\bf r} - \theta_{\bf r'} > \pi$ for a horizontal bond
$({\bf R,R'})$ or $\theta_{\bf r'} - \theta_{\bf r} > \pi$ for a
vertical bond $({\bf R,R'})$, and $({\bf R' \to R})$ in the
opposite case. The introduced representation is unique for a given
microstate of the spin system (the revers statement is not true,
of course).

The most basic structural unit that can be distinguished in the
representation we have built is a path $L$ (either straight or
steps-like) connecting two sites of the dual lattice, formed by
one or several bonds connected together so that their directions
comply with some general direction of the path. In the most
general case, that path can be either closed or not closed.

While a closed path $L$ represents a trivial situation, the spin
configuration with $L$ starting and ending at different sites of
the dual lattice is of great interest and is called a
vortex-antivortex pair (it can be said that the vortex and the
antivortex are centered at the ends of the path $L$). The above
concerns vortices with topological charges $\pm 1$; pairs of
vortices with higher values of topological charge can be defined
in terms of several paths that start at the vortex and end at the
antivortex. Paths that start at the same site but end at different
sites correspond to clusters of vortices with different absolute
values of charge (for example, $+2$ and $-1$, $-1$).

The regions around the vortex origins are characterized by
significant disorientation of spins and are called ``cores''.
Moderate spin-wave excitations, when $\theta_{\bf r} - \theta_{\bf
r'} \simeq 0, \pm 2\pi$ everywhere except for the vortex
``cores'', cannot destroy the vortex-antivortex pair unless the
two defects annihilate at the same point.

Short-range exchange forces between spins lead to long-range
effective interaction between vortices. The energy of this
interaction can be explicitly singled out in the Hamiltonian of
the Villain model, and turns out to depend only on the essentially
inherent characteristic of the vortices called \textit{topological
charge}~\cite{Villain75}. It will be shown that the logarithmic
asymptotic form, obtained by Villain for the attraction energy of
the vortex and the antivortex at large separations, in fact holds
sufficiently well on a lattice up to the smallest possible
separation of one lattice spacing $a$ (if neglecting the subtle
anisotropy effects).

On the contrary, the corresponding energy which we estimate for
the $2D$ $XY$ model turns out to deviate from the logarithmic law
at small separations. In particular, our result for the energy
needed to create a vortex-antivortex pair is approximately $6.6 J$
($J$ is the coupling constant), in contrast to $9.9 J$ of the
Villain model \cite{Villain75}, and in reasonable agreement with
the recent results of Monte Carlo
simulations~\cite{OtaOta95,GuptaBaillie92}.

The details of the results announced above can be found in Section
\ref{II}.

\subsection{\label{intro-B}Spin vacancies}

An aspect of the spin vortex behavior, which only recently drew
attention of the researchers, is the effective interaction with
nonmagnetic inclusions in the
lattice~\cite{Mol02,Wysin03,Pereira03}. Such spin vacancies are
part of the models with quenched
disorder~\cite{Leonel03,Wysin05a,Wysin05b,Berche03} and the
lattice gas spin models~\cite{ChamatiRomano06,ChamatiRomano07}.
Here, we will focus, however, not on the thermodynamic quantities,
but on the effective Hamiltonian which describes the interaction
between spin vortices and vacancies.

To our knowledge, the first theoretical works devoted to this
problem demonstrated global deformation of the vortex structure
caused by a single vacancy and repulsive interaction between the
vortex origin and the vacancy~\cite{Mol02,Leonel03}. This result
was essentially caused by an application of the KT continuum model
which required representation of the vacancy by a cutout of a
finite size in the continuous spin field. Subsequently, the same
authors denied this nonphysical result, on the basis of their spin
dynamics simulations~\cite{Pereira03}.

The problem was resolved phenomenologically, postulating that the
vacancy does not change the vortex structure (or the change is
negligible)~\cite{Pereira03}. Under this assumption, the KT theory
led to the attractive interaction which agreed with the results of
computer simulations. However, this approach, giving correct
qualitative picture, was not able to describe the particular
details of the lattice under consideration.

In our study, based on the Villain model, we obtain the effective
Hamiltonian describing interaction between spin vortices and spin
vacancies on a square lattice.

For example, as it will be shown in this paper, the interaction
energy for an individual vortex of topological charge $q$ at point
$\bf R$ and a spin vacancy at $\bf r$ reads:
\begin{equation}\label{E-intro}
E(|{\bf r-R}|) = - (\pi - 1) \frac{J q^2}{|{\bf r-R}|^2} +
O\left(|{\bf r-R}|^{-2}\right)\ ,
\end{equation}
i.e. the vacancy and the vortex attract each other.

Eq.~(\ref{E-intro}) is the asymptotic expression which in fact
holds well enough for separations as small as just a few lattice
spacings. It will be argued that in the $2D$ $XY$ model this
energy differs considerably from (\ref{E-intro}) for small
separations $|{\bf r-R}|$. For example, the vortex-on-vacancy
pinning energy of the Villain model $E(a) = -(3\pi-4) J q^2 \simeq
-5.425 Jq^2$, in contrast to that of the $2D$ $XY$ model observed
in spin dynamics simulations, $- 3.54 J$ \cite{Pereira03}, and
other numerical studies, $-3.178 J$ \cite{Wysin03}.

The details of the results announced here can be found in Section
\ref{III}.

\section{\label{II}Vortices in the Villain and $2D$ $XY$ models}

\subsection{\label{II-A}Topological charges in the Villain model}

Studying the low-temperature properties of the model
(\ref{H_XY_0}) it would be natural to apply the spin-wave
(harmonic) approximation (SWA), i.e. to replace $1 -
\cos(\theta_{\bf r} - \theta_{\bf r'})$ in (\ref{H_XY_0}) with
$\frac{1}{2}(\theta_{\bf r} - \theta_{\bf r'})^2$. Indeed, this
allows to examine many important properties of the low-temperature
phase of this
model~\cite{Wegner67,TobochnikChester79,Bramwell_et_al01}.
However, the states with $|\theta_{\bf r} - \theta_{\bf r'}| >
\pi$, which are crucial when considering spin vortices, will have
non-physical energy in this case. So, the proper harmonic
approximation must be
\begin{equation}\label{H_2DXY_harmonic}
H_{2DXY} \simeq \frac{J}{2} \sum_{\left<\bf r,r'\right>}
\left(\theta_{\bf r} - \theta_{\bf r'} - 2\pi m(\theta_{\bf
r}-\theta_{\bf r'})\right)^2
\end{equation}
with
\begin{equation}\nonumber
m(\theta_{\bf r}-\theta_{\bf r'}) = \Bigg\{ \begin{array}{lll}
+1,\ \theta_{\bf r}-\theta_{\bf r'} > \pi ;\\ -1,\  \theta_{\bf
r}-\theta_{\bf r'} < -\pi ;\\ 0,\ |\theta_{\bf r}-\theta_{\bf r'}|
< \pi .
\end{array}
\end{equation}
At low temperatures, $m(\theta_{\bf r}-\theta_{\bf r'})$ can be
considered as independent degrees of freedom taking discrete
values $0,\pm 1$. In turn, this leads to the Hamiltonian of the
Villain model:
\begin{equation}\label{H_Vill}
H = \frac{J}{2} \sum_{\left<\bf r,r'\right>} \left(\theta_{\bf r}
- \theta_{\bf r'} - 2\pi m_{\bf r,r'}\right)^2
\end{equation}
(obviously, $m_{\bf r,r'} = - m_{\bf r',r}$).

Assuming that
\begin{equation}\label{theta=phi+psi}
\theta_{\bf r} = \varphi_{\bf r} + \psi_{\bf r}\ ,
\end{equation}
where $\varphi_{\bf r}$ and $\psi_{\bf r}$ are chosen so that
$|\varphi_{\bf r} - \varphi_{\bf r'}| < \pi$ for any pair of spins
in the system, i.e one can say that the field $\varphi_{\bf r}$ is
vortexless, and all the vortices are ``contained'' in $\psi_{\bf
r}$, the Hamiltonian (\ref{H_Vill}) can be written as:
\begin{eqnarray}\nonumber
H &=& \frac{J}{2} \sum_{\left<{\bf r, r'}\right>}
\left[(\varphi_{\bf r} - \varphi_{\bf r'})^2 + (\psi_{\bf r} -
\psi_{\bf r'} - 2\pi m_{\bf r,r'})^2\right]\\\label{H_Vill_1} &&+\
J \sum_{\bf r}\varphi_{\bf r}\sum_{\bf u}\left(\psi_{\bf r} -
\psi_{\bf r+u} - 2\pi m_{\bf r,r+u}\right)\qquad
\end{eqnarray}
with ${\bf u} = (\pm a,0), (0,\pm a)$ and lattice spacing $a$.

\begin{figure}[h]

\includegraphics[width=0.12\textwidth]{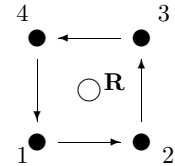}

\caption{Plaquette of sites $1,2,3,4$ of the initial lattice
adjacent to site ${\bf R}$ of the dual lattice.}\label{Fig2}
\end{figure}

Following Villain \cite{Villain75}, one can choose $\psi_{\bf
r}(\{m_{\bf r,r'}\})$ such that $\varphi_{\bf r}$ and $\psi_{\bf
r}$ decouple in the Hamiltonian, i.e. the last term in
(\ref{H_Vill_1}) vanishes:
\begin{equation}\label{H_phipsi=0_condition}
\sum_{\bf u}\left(\psi_{\bf r} - \psi_{\bf r+u} - 2\pi m_{\bf
r,r+u}\right)\ =\ 0\ \textrm{ for all }\ {\bf r}\ .
\end{equation}
This is realized when
\begin{eqnarray}\label{psi}
\psi_{\bf r} &=& \frac{\pi}{2}\sum_{\bf R}\Big\{\
(m_{3,4}-m_{1,2})I_{sc}(x-X,y-Y)\\\nonumber &&+\
(m_{4,1}-m_{2,3})I_{sc}(y-Y,x-X) + (m_{1,2}\\\nonumber && -\
m_{2,3} + m_{3,4}-m_{4,1})I_{ss}(x-X,y-Y)\ \Big\}\phantom{^\big|}
\end{eqnarray}
(see Fig. \ref{Fig2}) where $\bf R$ are sites of the dual lattice,
which are situated in the centers of elementary cells of the
original lattice, and functions $I_{sc}$ and $I_{ss}$ are given by
(\ref{Isc_def}) and (\ref{Iss_def}). (The asymptotic properties of
$I_{sc}$ and $I_{ss}$ are analyzed in Appendix \ref{appendxA}.) In
fact, Eq.~(\ref{psi}) is another way of presenting the expression
obtained by Villain~\cite{Villain75}.

In (\ref{psi}), the sum over ${\bf R}=(X,Y)$ spans the sites of
the dual lattice, while coordinate $\bf r$ represents a site of
the original lattice, therefore, $X-x$ and $Y-y$ can be always
presented as $(2n-1)\frac{a}{2}$, where $n$ is an integer. The
short notation
\begin{equation}\nonumber \textstyle
I_{sc(ss)}\left((2n-1)\frac{a}{2},(2m-1)\frac{a}{2}\right) \equiv
I^{nm}_{sc(ss)}
\end{equation}
will be helpful.

Due to the properties: $I_{sc}(-X,Y) = -I_{sc}(X,Y)$,
$I_{sc}(X,-Y) = I_{sc}(X,Y)$, $I_{ss}(-X,Y) = -I_{ss}(X,Y)$,
$I_{ss}(X,Y) = I_{ss}(Y,X)$, it is enough to define $I^{nm}_{sc}$
and $I^{nm}_{ss}$ only for $n,m$ being positive nonzero integers
(natural numbers), thus they can be presented as infinite
matrices. In the thermodynamic limit, one has (see Appendix A for
the general expression) $I^{nm}_{sc}$
\begin{equation}\label{I^sc_nm}
= \left(
\begin{array}{ccccc}
\frac{1}{\pi} & \frac{1}{2}-\frac{1}{\pi} & \frac{3}{2}-\frac{13}{3\pi} & \frac{11}{2}-\frac{17}{\pi} & \ldots \\
&&&&\\
-\frac{3}{2}+\frac{5}{\pi} & \frac{1}{3\pi} & -\frac{3}{2}+\frac{5}{\pi} & -\frac{15}{2}+\frac{119}{5\pi} & \ldots \\
&&&&\\
-\frac{15}{2}+\frac{71}{3\pi} & \frac{5}{2}-\frac{23}{3\pi} & \frac{1}{5\pi} & \frac{5}{2}-\frac{23}{3\pi} & \ldots \\
&&&&\\
-\frac{77}{2}+\frac{121}{\pi} & \frac{35}{2}-\frac{823}{15\pi} & -\frac{7}{2}+\frac{167}{15\pi} & \frac{1}{7\pi} & \ldots \\
\vdots & \vdots & \vdots & \vdots & \ddots
\end{array} \right)
\end{equation}
and $I^{nm}_{ss}$
\begin{equation}\label{I^ss_nm}
= \left(
\begin{array}{ccccc}
\frac{1}{2}-\frac{1}{\pi} & 1-\frac{3}{\pi} & 5-\frac{47}{3\pi} & 26-\frac{245}{3\pi} & \ldots \\
&&&&\\
1-\frac{3}{\pi} & -\frac{1}{2}+\frac{5}{3\pi} & -2+\frac{19}{3\pi} & -13+\frac{613}{15\pi} & \ldots \\
&&&&\\
5-\frac{47}{3\pi} & -2+\frac{19}{3\pi} & \frac{1}{2}-\frac{23}{15\pi} & 3-\frac{47}{5\pi} & \ldots \\
&&&&\\
26-\frac{245}{3\pi} & \ -13+\frac{613}{15\pi} & \ 3-\frac{47}{5\pi} & \ -\frac{1}{2}+\frac{167}{105\pi} & \ldots \\
\vdots & \vdots & \vdots & \vdots & \ddots
\end{array} \right) .
\end{equation}

Note that there was no reason for presenting $I^{nm}_{sc}$,
$I^{nm}_{ss}$ as matrices, other than the convenient
visualization.

We have found that the exact values of $\psi_{\bf r}$, provided by
(\ref{psi}), (\ref{I^sc_nm}), (\ref{I^ss_nm}) within the vortex
core, are quite close to that of its asymptotic form found by
Villain~\cite{Villain75}:
\begin{equation}\label{vort_field}
\psi_{\bf r} \simeq \sum_{\bf R} q_{\bf R} \Phi_{\bf r}({\bf R})\
,
\end{equation}
where $\Phi_{\bf r}({\bf R})$ is the polar coordinate of point
$\bf r$ in the coordinate system with its origin at point $\bf R$
(the reference angles are such that $\psi_{\bf r} - \psi_{\bf r'}
> \pi$ if $m_{\bf r,r'} = 1$, $\psi_{\bf r} - \psi_{\bf r'}
< -\pi$ if $m_{\bf r,r'} = - 1$, and $|\psi_{\bf r} - \psi_{\bf
r'}| < \pi$ if $m_{\bf r,r'} = 0$), and
\begin{equation}\label{q_on_plaquette}
q_{\bf R} = m_{1,2} + m_{2,3} + m_{3,4} + m_{4,1}
\end{equation}
is the topological charge defined at site $\bf R$ of the dual
lattice (see Fig. \ref{Fig2}).


\begin{figure}[h]
\includegraphics[width=0.4\textwidth]{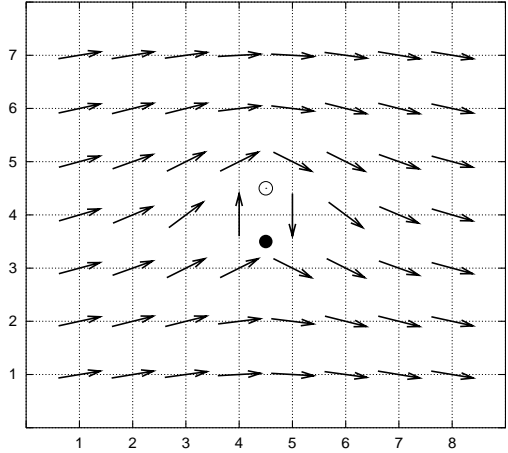}
\caption{Field $\psi_{(i,j)}$ for a vortex-antivortex pair
situated at sites (4.5.4.5) and (4.5,3.5) (the open and filled
circles represent the vortex and the antivortex, respectively)
given by Eq.~(\ref{vort_field}) (Tab.~\ref{Tab1}). The difference
with the exact result following from (\ref{psi}) (Tab.~\ref{Tab2})
is insignificant within the resolution of the present picture.}
\label{Fig3}
\end{figure}

Compare, for example, the field $\psi_{\bf r}$ given by Eqs.
(\ref{psi}) and (\ref{vort_field}) for a vortex-antivortex pair
with the minimal separation (see Fig. \ref{Fig3}), shown in Tabs.
\ref{Tab1} and \ref{Tab2}. .


It is worth mentioning that Eq.~(\ref{vort_field}) can be derived
from (\ref{psi}), using the asymptotic form of $I_{sc}$ and
$I_{sc}$, Eqs. (\ref{Isc_asympt}), (\ref{Iss_asympt}), and
integrating (instead of summing over $\bf R$) along a properly
chosen path $L$ (or along several paths for the vortices with
higher topological charges, see Section IA) connecting the vortex
with its antivortex. We have verified that the form of the field
$\psi_{\bf r}$ given by Eq. (\ref{psi}) is independent of the
particular form of this path $L$.


\begin{table}[h]
\caption{\label{Tab1} Field $\psi_{(i,j)}$ (see Fig.~\ref{Fig3})
given by Eq.~(\ref{vort_field})}
\begin{ruledtabular}
\begin{tabular}{ccccc}
\qquad\quad \vline & $i=1$ & $i=2$ & $i=3$ & $i=4$ \\
\hline $j=1$\ \ \vline & $0.1651$ & $0.1651$
& $0.1355$ & $0.0555$ \\
$j=2$\ \ \vline & $0.2154$ & $0.2450$ &
$0.2450$ & $0.1244$ \\
$j=3$\ \ \vline & $0.2630$ & 0.3430 & 0.4636 & 0.4636 \\
$j=4$\ \ \vline & 0.2838 & 0.3948 & 0.6435 & $\pi/2$\\
\end{tabular}
\end{ruledtabular}
\end{table}




\begin{table}[h]
\caption{\label{Tab2} Field $\psi_{(i,j)}$ (see Fig.~\ref{Fig3})
given by Eq.~(\ref{psi})}
\begin{ruledtabular}
\begin{tabular}{ccccc}
\qquad\quad \vline & $i=1$ & $i=2$ & $i=3$ & $i=4$ \\
\hline \qquad\quad \vline
\\
$j=1$\ \ \vline & $\begin{array}{ll}
\frac{26}{15}-\frac{\pi}{2} \\ \simeq 0.1625 \end{array}$ &
$\begin{array}{ll} \frac{26}{15}-\frac{\pi}{2} \\ \simeq 0.1625
\end{array}$ & $\begin{array}{ll} \frac{9}{2}\pi - 14 \\ \simeq 0.1372
\end{array}$ & $\begin{array}{ll} \frac{118}{5}-\frac{25}{2}\pi \\ \simeq 0.0635
\end{array}$ \\ \qquad\quad \vline
\\
$j=2$\ \ \vline & $\begin{array}{ll}
\frac{11}{2}\pi - \frac{256}{15} \\ \simeq 0.2121 \end{array}$ &
$\begin{array}{ll} \frac{\pi}{2} - \frac{4}{3} \\ \simeq 0.2375
\end{array}$ & $\begin{array}{ll} \frac{\pi}{2} - \frac{4}{3} \\
\simeq 0.2375 \end{array}$ & $\begin{array}{ll} 8 -
\frac{5}{2}\pi \\ \simeq 0.1460 \end{array}$ \\ \qquad\quad \vline
\\
$j=3$\ \ \vline & $\begin{array}{ll} \frac{194}{3}-\frac{41}{2}\pi
\\ \simeq 0.2641 \end{array}$ & $\begin{array}{ll} \frac{34}{3}-\frac{7}{2}\pi
\\ \simeq 0.3378 \end{array}$ & $\begin{array}{ll} 2 -
\frac{\pi}{2} \\ \simeq 0.4292 \end{array}$ & $\begin{array}{ll} 2
- \frac{\pi}{2} \\ \simeq 0.4292 \end{array}$ \\ \qquad\quad
\vline
\\
$j=4$\ \ \vline & $\begin{array}{ll} \frac{63}{2}\pi -
\frac{296}{3} \\ \simeq 0.2934 \end{array}$ & $\begin{array}{ll}
\frac{13}{2}\pi - 20 \\ \simeq 0.4203 \end{array}$ &
$\begin{array}{ll} \frac{3}{2}\pi - 4 \\ \simeq 0.7124 \end{array}$ & $\pi/2$\\
\end{tabular}
\end{ruledtabular}
\end{table}


\subsection{\label{II-B}Interaction between vortices in the Villain and $2D$ $XY$ models}

If $\psi_{\bf r}$ is given by Eq.~(\ref{psi}), the
Hamiltonian~(\ref{H_Vill_1}) can be reduced to
\begin{equation}\label{H_Vill_2}
H = \frac{J}{2}\sum_{\left<{\bf r, r'}\right>}(\varphi_{\bf r} -
\varphi_{\bf r'})^2 + \sum_{\bf R, R'} q_{\bf R}q_{\bf R'}V({\bf
R-R'})\ ,
\end{equation}
where topological charge $q_{\bf r}$ is defined by
Eq.~(\ref{q_on_plaquette}). Now the vortex interaction energy is
given by the second term in the Hamiltonian (\ref{H_Vill_2}) with
\begin{equation}\label{V(R-R')_def}
V({\bf R-R'})\ =\ \frac{\pi^2 J}{N} \sum_{\bf k} \frac{\cos
k_x(X-X') \cos
k_y(Y-Y')}{\sin^2\frac{k_xa}{2}+\sin^2\frac{k_ya}{2}}\ .
\end{equation}

In the thermodynamic limit one can replace the sum over the first
Brillouin zone in (\ref{V(R-R')_def}) with an integral, and then,
since the difference between the Cartesian coordinates of the
vortices centered on sites of the dual lattice is always an
integer number of lattice spacing $a$: $X-X' = na$, $Y-Y' = ma$,
following the same scheme of integration which was applied in
Appendix A to obtain (\ref{Isc_exact}), (\ref{Iss_exact}), one
has:
\begin{eqnarray}\nonumber
V(na,ma) = \sum_{i=0}^{n} \frac{(-1)^i(2n)!}{(2(n-i))!(2i)!}
\sum_{j=0}^{m} \frac{(-1)^j(2m)!}{(2(m-j))!(2j)!}\\\nonumber
\times \sum_{k=0}^{n-i} \frac{(-1)^k(n-i)!}{(n-i-k)!\ k!}
\sum_{l=0}^{m-j} \frac{(-1)^l(m-j)!}{(m-j-l)!\ l!}
F(i+k,j+l)\\\label{V(R-R')_exact}
\end{eqnarray}
with $F(p,q)$ given by Eq.~(\ref{F(p,q)}).

\begin{figure}[h]
\includegraphics[angle=-90,width=0.48\textwidth]{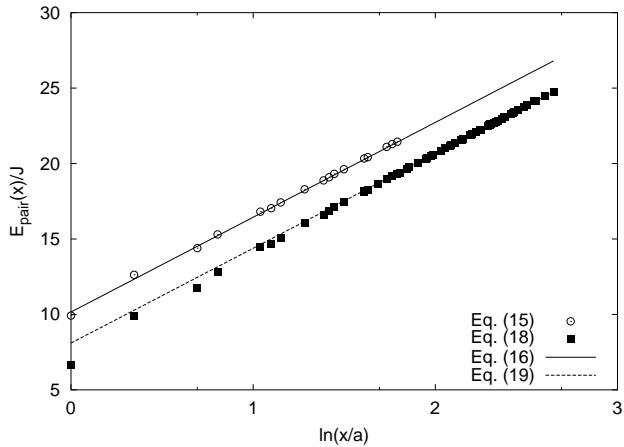}
\caption{The energy of a pair of topological charges $q_{\bf
R}=+1$, $q_{\bf R'}=-1$ in the Villain and $2D$ $XY$ models as a
function of the separation $x = |{\bf R-R'}|$. Open circles
represent the exact result for the Villain model, following from
(\ref{V(R-R')_exact}); filled squares represent the numerical
result for the $2D$ $XY$ model (see (\ref{E_XY(R)})); solid and
dashed lines are the asymptotic forms (\ref{E^vill_asympt}) and
(\ref{E^2dxy_asympt}).}\label{Fig6}
\end{figure}

Then the energy of a vortex-antivortex pair, $q_{\bf R} = +1$ and
$q_{\bf R'} = -1$, which follows from (\ref{H_Vill_2}), is
$E_{\mathrm{pair}}(x) = - V(x)$, where $x = |{\bf R-R'}|$ is the
distance between the vortex and the antivortex. Comparing
$E_{\mathrm{pair}}(x)$ that follows from (\ref{V(R-R')_exact})
with the asymptotic expression found by Villain~\cite{Villain75}:
\begin{equation}\label{E^vill_asympt}
E_{\mathrm{pair}}(x) \simeq 10.158 J + 2\pi J\ln(x/a),
\end{equation}
see Fig.~\ref{Fig6}, we notice a fine agreement.  The low number
of points for small $x/a$ is due to limited number of
possibilities to situate the pair on a lattice, and the
``oscillation'' of data is the anisotropy effect for different
orientations of vector ${\bf R-R'}$.


As it was already mentioned in the Introduction, the cores of
vortices are characterized by large angles between the neighboring
spins, so the harmonic approximation (\ref{H_2DXY_harmonic})
cannot give the correct value of the energy of vortex cores in the
model (\ref{H_XY_0}). Obviously, this can lead to different
inter-vortex interaction energies in the Villain and $2D$ $XY$
models.

To estimate the energy of the vortex-antivortex interaction in the
$2D$ $XY$ model, $E^{2DXY}_{\mathrm{pair}}(x)$, we consider field
\begin{equation}\label{vort_field_pair}
\psi'_{\bf r} = \Phi_{\bf r}({\bf R}) - \Phi_{\bf r}({\bf R}')
\end{equation}
(see Eq. (\ref{vort_field})), which corresponds to the topological
charges $q_{{\bf R}} = +1$ and $q_{{\bf R}'} = -1$, and assume
that
\begin{equation}\label{E_XY(R)}
E^{2DXY}_{\mathrm{pair}}(x) =  J \sum_{\left<{\bf r, r'}\right>}
\left[1 - \cos(\psi'_{\bf r} - \psi'_{\bf r'})\right] ,
\end{equation}
performing the summation numerically over a system of sufficiently
large size. We are aware that this assumption is not grounded,
since $\varphi_{\bf r}$ and $\psi_{\bf r}$ cannot be decoupled in
the Hamiltonian (\ref{H_XY_0}), but it may be instructive.

The quantity which is accessible for measurement in Monte Carlo
simulations is the vortex-antivortex pair creation energy in the
$2D$ $XY$ model, i.e. the energy of a vortex and its antivortex at
the minimal separation $a$ (see Fig. \ref{Fig3}):
$E^{2DXY}_{\mathrm{pair}}(a)$. The microcanonical Monte Carlo
simulations showed that $E^{2DXY}_{\mathrm{pair}}(a) \simeq 7.3 J$
\cite{OtaOta95}, while the canonical MC simulations gave $7.55 J$
\cite{GuptaBaillie92}. Our estimation which follows from
(\ref{E_XY(R)}) is $E^{2DXY}_{\mathrm{pair}}(a) \simeq 6.6 J$, in
reasonable agreement with the mentioned computer experiments (the
exact result for the Villain model is $\pi^2 J \simeq 9.9 J$, see
(\ref{V(R-R')_exact})).

Comparing the result of (\ref{E_XY(R)}) to the vortex-antivortex
interaction energy in the Villain model, see Fig. \ref{Fig6}, we
see that while at large separations
\begin{equation}\label{E^2dxy_asympt}
E^{2DXY}_{\mathrm{pair}}(x)\ \simeq\ 8.1 J\ +\ 2\pi J\ln(x/a),
\end{equation}
$E^{2DXY}_{\mathrm{pair}}(x)$ deviates considerably from the
logarithmic form as the vortex and its antivortex approach each
other.

\section{\label{III}Interaction between vortices and spin vacancies}

\subsection{\label{III-A}Hamiltonian of the Villain model with spin
vacancies}

With the use of variables $c_{\bf r}$, taking values 1 and 0
depending on whether site $\bf r$ is occupied with a spin or
``empty'', respectively, the Hamiltonian of the Villain model with
spin vacancies can be presented as \cite{Kapikranian08}:
\begin{equation}\label{H_vill_dil}
H = \frac{J}{2}\sum_{\left<{\bf r,r'}\right>} (\theta_{\bf r} -
\theta_{\bf r'} - 2\pi m_{\bf r,r'})^2 c_{\bf r} c_{\bf r'} .
\end{equation}
Alternatively, it can be written via variables $p_{\bf r} = 1 -
c_{\bf r}$ as
\begin{equation}\label{H=H0+H1+H2}
H = H_0 + \sum_{\bf r} p_{\bf r} H_1({\bf r}) + \sum_{\left<{\bf
r,r'}\right>} p_{\bf r} p_{\bf r'}  H_2({\bf r,r'})\ ,
\end{equation}
where $H_0$ is the Hamiltonian of the Villain model without
vacancies (Eq.~(\ref{H_vill_dil}) with all $c_{\bf r} = 1$),
\begin{equation}\label{H_1(r)}
H_1({\bf r}) = - \frac{J}{2} \sum_{\bf u} (\theta_{\bf r} -
\theta_{\bf r+u} - 2\pi m_{\bf r,r+u})^2
\end{equation}
with ${\bf u} = (\pm a,0), (0,\pm a)$ is the change in energy
caused by the removal of the four bonds adjacent to the spinless
site ${\bf r}$, and
\begin{equation}
H_2({\bf r,r'}) = \frac{J}{2} (\theta_{\bf r} - \theta_{\bf r'} -
2\pi m_{\bf r,r'})^2
\end{equation}
compensates the double removal of a common bond of two neighboring
sites ${\bf r}$ and ${\bf r}'$ when there happen vacancies on
neighboring sites.

Applying (\ref{theta=phi+psi}), one can distinguish in the
Hamiltonian (\ref{H_vill_dil}) terms dependent on vortexless field
$\varphi_{\bf r}$ and/or vortex field $\psi_{\bf r}$ (marking them
with indices $\varphi$ and $\psi$):
\begin{equation}\label{H_phi+H_psi+H_phipsi}
H = H^{\varphi} + H^{\psi} + H^{\varphi,\psi} .
\end{equation}
The first term, considered separately, describes a system of
planar spins with angles $\varphi_{\bf r}$ on a diluted lattice in
the SWA, which was the subject of studies \cite{Berche03} and
\cite{Kapikranian07}, for example. Here, we focus primarily on the
last two terms that are connected to the presence of vortices in
the system.

Notice that taking $\psi_{\bf r}$ in the form of Eq.~(\ref{psi})
does not lead to decoupling of $\varphi_{\bf r}$ and $\psi_{\bf
r}$ in the Villain model with spin vacancies
($H^{\varphi,\psi}\neq 0$).

\subsection{\label{III-B}Hamiltonian of the Villain model with spin vacancies
            in the Fourier-transformed variables}

Fourier transformation of variables $\varphi_{\bf r}$, $\psi_{\bf
r}$ and $m_{{\bf r},{\bf r}'}$ allows to manipulate Hamiltonian
(\ref{H_vill_dil}) with much ease. The corresponding Fourier
transforms $\varphi_{\bf k}$, $\psi_{\bf k}$ and  $m^\alpha_{\bf
k}$ ($\alpha = x,y$ stands to distinguish two sets of Fourier
transforms that correspond to ``vertical/horizontal'' orientation
of bond $({\bf r,r'})$) can be introduced via the following
relations:
\begin{eqnarray}\label{phi_Fourier}
\varphi_{\bf k} &=& \frac{1}{\sqrt{N}} \sum_{\bf r} e^{i{\bf kr}}
\varphi_{\bf r},\quad \varphi_{\bf r} = \frac{1}{\sqrt{N}}
\sum_{\bf k} e^{-i{\bf kr}} \varphi_{\bf k},\quad
\\\label{psi_Fourier}
\psi_{\bf k} &=& \frac{1}{\sqrt{N}} \sum_{\bf r} e^{i{\bf kr}}
\psi_{\bf r},\quad \psi_{\bf r} = \frac{1}{\sqrt{N}} \sum_{\bf k}
e^{-i{\bf kr}} \psi_{\bf k} ,
\end{eqnarray}
\begin{eqnarray}\label{m_Fourier}
m^\alpha_{\bf k} &=& \frac{1}{\sqrt{N}} \sum_{\bf r}
e^{i\left({\bf kr} + \frac{k_\alpha a}{2}\right)} m_{{\bf r},{\bf
r} + {\bf u}_\alpha} ,
\\\nonumber m_{{\bf r},{\bf
r} + {\bf u}_\alpha} &=& \frac{1}{\sqrt{N}} \sum_{\bf k}
e^{-i\left({\bf kr} + \frac{k_\alpha a}{2}\right)} m^\alpha_{\bf
k}\ ,\quad \alpha=x,y\ ,
\end{eqnarray}
where $N$ is the number of sites in the lattice and the sums over
$\bf r$ and $\bf k$ span the original lattice and the 1st
Brillouin zone of the reciprocal lattice, respectively. Note that
in (\ref{m_Fourier}) ${\bf u}_x = (a,0)$, ${\bf u}_y = (0,a)$, so
the property $m_{\bf r,r'} = - m_{\bf r',r}$ is supposed to be
used to obtain the Fourier transform of $m_{{\bf r} + {\bf
u}_\alpha,{\bf r}}$.

Then, for the field $\psi_{\bf r}$ given by Eq.~(\ref{psi}) one
has the Fourier transform \cite{Villain75}
\begin{equation}\label{psi_fourier}
\psi_{\bf k} = - i\pi \frac{m^x_{\bf k} \sin\frac{k_xa}{2} +
m^y_{\bf k} \sin\frac{k_ya}{2}}{\sin^2\frac{k_xa}{2} +
\sin^2\frac{k_ya}{2}} .
\end{equation}
Now, using (\ref{psi_fourier}), the condition
(\ref{H_phipsi=0_condition}) can be easily checked.

After applying (\ref{m_Fourier}) and introducing the
Fourier-trans\-form of the topological charge $q_{\bf r}$:
\begin{equation}\label{q_Fourier}
q_{\bf k} = \frac{1}{\sqrt{N}} \sum_{\bf R} e^{i{\bf kR}} q_{\bf
R},\quad q_{\bf R} = \frac{1}{\sqrt{N}} \sum_{\bf k} e^{-i{\bf
kR}} q_{\bf k}\ ,
\end{equation}
Eq.~(\ref{q_on_plaquette}) takes the form
\begin{equation}
q_{\bf k} = 2i \left( m^x_{\bf k} \sin\frac{k_ya}{2} - m^y_{\bf k}
\sin\frac{k_xa}{2} \right)\ .
\end{equation}

Then, it is quite straightforward to obtain (the reader is
referred to Eqs. (\ref{H=H0+H1+H2}) and
(\ref{H_phi+H_psi+H_phipsi}) to understand the upper and bottom
indices in the left sides of the equations)
\begin{widetext}
\begin{equation}\label{H_1^phi-psi_fourier}
H_{1}^{\varphi\psi}({\bf r}) = \frac{4\pi J}{N} \sum_{\bf k}
\sum_{\bf k'} \varphi_{\bf k} q_{\bf k'}
\frac{\cos{\textstyle\frac{(k_x+k'_x)a}{2}} \sin\frac{k_xa}{2}
\sin\frac{k'_ya}{2} - \cos{\textstyle\frac{(k_y+k'_y)a}{2}}
\sin\frac{k_ya}{2} \sin\frac{k'_xa}{2}}{\sin^2\frac{k'_xa}{2} +
\sin^2\frac{k'_ya}{2}}\ e^{-i({\bf k+k'}){\bf r}}\ ,
\end{equation}
\begin{equation}\label{H_1^psi-psi_fourier}
H_{1}^{\psi}({\bf r}) = \frac{\pi^2 J}{N} \sum_{\bf k} \sum_{\bf
k'} q_{\bf k} q_{\bf k'}
\frac{\cos{\textstyle\frac{(k_x+k'_x)a}{2}} \sin\frac{k_ya}{2}
\sin\frac{k'_ya}{2} - \cos{\textstyle\frac{(k_y+k'_y)a}{2}}
\sin\frac{k_xa}{2} \sin\frac{k'_xa}{2}}{\left(\sin^2\frac{k_xa}{2}
+ \sin^2\frac{k_ya}{2}\right)\left(\sin^2\frac{k'_xa}{2} +
\sin^2\frac{k'_ya}{2}\right)}\ e^{-i({\bf k+k'}){\bf r}}\ ,
\end{equation}
\begin{eqnarray}\nonumber
H_{2}^{\varphi\psi}({\bf r,r'}) &=& -\frac{2\pi J}{N} \sum_{\bf k}
\sum_{\bf k'} \varphi_{\bf k} q_{\bf k'} e^{-i({\bf k+k'}){\bf r}}
\Bigg( \left( \delta_{{\bf r'-r},{\bf u}_x}\
e^{-i\frac{(k_x+k'_x)a}{2}} + \delta_{{\bf r'-r},-{\bf u}_x}\
e^{i\frac{(k_x+k'_x)a}{2}} \right)\textstyle \sin\frac{k_xa}{2}
\sin\frac{k'_ya}{2}
\\\label{H_2^phi-psi_fourier}
&&- \left( \delta_{{\bf r'-r},{\bf u}_y}\
e^{-i\frac{(k_y+k'_y)a}{2}} + \delta_{{\bf r'-r},-{\bf u}_y}\
e^{i\frac{(k_y+k'_y)a}{2}} \right)\textstyle \sin\frac{k_ya}{2}
\sin\frac{k'_xa}{2}\Bigg) \left(\sin^2\frac{k'_xa}{2} +
\sin^2\frac{k'_ya}{2}\right)^{-1}\ ,
\end{eqnarray}
\begin{eqnarray}\nonumber
H_{2}^{\psi}({\bf r,r'}) &=& -\frac{\pi^2 J}{2N} \sum_{\bf k}
\sum_{\bf k'} q_{\bf k} q_{\bf k'} e^{-i({\bf k+k'}){\bf r}}
\Bigg( \left( \delta_{{\bf r'-r},{\bf u}_x}\
e^{-i\frac{(k_x+k'_x)a}{2}} + \delta_{{\bf r'-r},-{\bf u}_x}\
e^{i\frac{(k_x+k'_x)a}{2}} \right)\textstyle \sin\frac{k_ya}{2}
\sin\frac{k'_ya}{2}
\\\label{H_2^psi-psi_fourier}
&&- \left( \delta_{{\bf r'-r},{\bf u}_y}\
e^{-i\frac{(k_y+k'_y)a}{2}} + \delta_{{\bf r'-r},-{\bf u}_y}\
e^{i\frac{(k_y+k'_y)a}{2}} \right)\textstyle \sin\frac{k_xa}{2}
\sin\frac{k'_xa}{2}\Bigg) \Big/{\displaystyle
\sum_{\alpha=x,y}}\sin^2\frac{k_\alpha a}{2} {\displaystyle
\sum_{\alpha=x,y}}\sin^2\frac{k'_\alpha a}{2}\ .\qquad
\end{eqnarray}

\subsection{\label{III-C}Attractive interaction between spin vortices and a spin vacancy}

Returning to variables $\varphi_{\bf r}$ and $q_{\bf R}$ in
Eqs.~(\ref{H_1^phi-psi_fourier}), (\ref{H_1^psi-psi_fourier}), one
has
\begin{eqnarray}\nonumber
H_{1}^{\varphi\psi}({\bf r}) &=& \pi J \sum_{\bf R} q_{\bf R}
\big\{\left(\varphi_{{\bf r}+{\bf u}_x} - \varphi_{{\bf r}-{\bf
u}_x}\right) I_{sc}(y-Y,x-X) - \left(\varphi_{{\bf r}+{\bf u}_y} -
\varphi_{{\bf r}-{\bf u}_y}\right) I_{sc}(x-X,y-Y)
\\\label{H_1^phi-psi} && + \big(\varphi_{{\bf r}+{\bf u}_y} +
\varphi_{{\bf r}-{\bf u}_y} - \varphi_{{\bf r}+{\bf u}_x} -
\varphi_{{\bf r}-{\bf u}_x}\big) I_{ss}(x-X,y-Y) \big\}
\end{eqnarray}
and
\begin{eqnarray}\nonumber
H_{1}^{\psi}({\bf r}) &=& -\ \pi^2 J \sum_{\bf R}\sum_{\bf R'}
q_{\bf R} q_{\bf R'}
\big\{I_{\mathrm{sc}}(x-X,y-Y)I_{\mathrm{sc}}(x-X',y-Y') +
I_{\mathrm{sc}}(y-Y,x-X) I_{\mathrm{sc}}(y-Y',x-X')
\\\label{H_1^psi-psi}
&& +\ 2I_{\mathrm{ss}}(x-X,y-Y)I_{\mathrm{ss}}(x-X',y-Y')\big\}
,\qquad
\end{eqnarray}
\end{widetext}
where $I_{sc}$ and $I_{ss}$ are defined by Eqs.~(\ref{Isc_def})
and (\ref{Iss_def}). Analogous expressions for
(\ref{H_2^phi-psi_fourier}) and (\ref{H_2^psi-psi_fourier}) can be
obtained easily.

In order to obtain the effective Hamiltonian describing
interaction between vacancies and topological charges only, one
has to integrate out $\varphi_{\bf r}$ in the partition function
\begin{equation}\label{Z_0}
Z = \mathrm{Tr}_{\varphi,\psi} e^{-\beta \left(H^{\varphi} +
H^{\psi} + H^{\varphi,\psi}\right)},
\end{equation}
so that
\begin{equation}\label{Z_1}
Z = \mathrm{Tr}_{\psi} e^{-\beta H^{\psi}_{\mathrm{eff}}},
\end{equation}
where $H^{\psi}_{\mathrm{eff}}$ is the desired Hamiltonian.

We have to restrict our consideration to the case of one spin
vacancy at site ${\bf r}^*$ to be able to use the results of
Appendix \ref{appendxB}. Then, using (\ref{H_eff}) and
(\ref{H_1^phi-psi_fourier}), one has the effective Hamiltonian
\begin{widetext}
\begin{eqnarray}\nonumber
H^{\psi}_{\mathrm{eff}}({\bf r}^*) &=& H_{1}^{\varphi\psi}({\bf
r}^*) + \frac{\pi^2 J}{N} \sum_{\bf k,k'} q_{\bf k} q_{\bf k'}
\textstyle \Big[ (\pi-2) \Big(\sin\frac{k_xa}{2}
\cos\frac{k_ya}{2} \sin\frac{k'_xa}{2} \cos\frac{k'_ya}{2}
{\textstyle +\ \sin\frac{k_ya}{2} \cos\frac{k_xa}{2}
\sin\frac{k'_ya}{2} \cos\frac{k'_xa}{2}} \Big)
\\\label{H_eff^psi_0} &&-\ 2\frac{4-\pi}{\pi-2}\ {\textstyle
\sin\frac{k_xa}{2} \sin\frac{k_ya}{2} \sin\frac{k'_xa}{2}
\sin\frac{k'_ya}{2}} \Big] e^{-i({\bf k+k'}){\bf r}^*}\Big/
\sum_{\alpha=x,y} {\textstyle \sin^2\frac{k_\alpha a}{2}}
\sum_{\alpha=x,y} {\textstyle \sin^2\frac{k'_\alpha a}{2}}.
\end{eqnarray}
Finally, using (\ref{H_1^psi-psi}) and (\ref{q_Fourier}), one can
write:
\begin{eqnarray}\nonumber
H^{\psi}_{\mathrm{eff}}({\bf r}^*) &=& -\ \pi^2 J \sum_{\bf
R}\sum_{\bf R'} q_{\bf R} q_{\bf R'} \Big\{ (\pi-1) \big[
I_{\mathrm{sc}}(x^*-X,y^*-Y)I_{\mathrm{sc}}(x^*-X',y^*-Y')
\\\label{H_eff^psi}
&&+\ I_{\mathrm{sc}}(y^*-Y,x^*-X) I_{\mathrm{sc}}(y^*-Y',x^*-X')
\big] + \frac{4}{\pi-2}\
I_{\mathrm{ss}}(x^*-X,y^*-Y)I_{\mathrm{ss}}(x^*-X',y^*-Y')\Big\} .
\end{eqnarray}
\end{widetext}


\begin{figure}[h]
\includegraphics[angle=-90,width=0.48\textwidth]{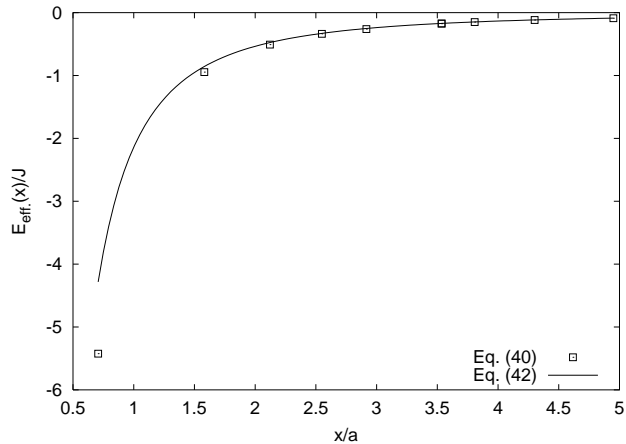}
\caption{Interaction energy of a vortex of charge $\pm 1$ and a
vacancy as a function of their separation $x$. Open squares
represent the exact result (\ref{H_eff^psi}) and the solid curve
represents the asymptotic expression
(\ref{E_asympt}).}\label{Fig8}
\end{figure}


While (\ref{Isc_exact}) and (\ref{Iss_exact}) provide the exact
value of (\ref{H_eff^psi}) for a discrete lattice, it is
instructive to find its asymptotic form:
\begin{eqnarray}\nonumber
H^{\psi}_{\mathrm{eff}}({\bf r}^*)\ =\ - (\pi - 1) Ja^2 \sum_{\bf
R,R'} q_{\bf R} q_{\bf R'} \qquad\qquad\qquad\quad
\\\nonumber \times\left[ \frac{({\bf r^* - R})({\bf
r^* - R'})}{|{\bf r^* - R}|^2 |{\bf r^* - R'}|^2}+
O\left(\frac{1}{|{\bf r^* - R}|^2 |{\bf r^* - R'}|^2}\right)
\right] ,\\\label{V_1^psi_asympt}
\end{eqnarray}
which follows from (\ref{Isc_asympt}) and (\ref{Iss_asympt}). If,
for example, one has a vortex of topological charge either $+$ or
$-1$ and a spin vacancy, separated by distance $x$,
(\ref{V_1^psi_asympt}) gives the energy of their interaction:
\begin{equation}\label{E_asympt}
E(x) = -J(\pi - 1)a^2/x^2 + O(1/x^2)
\end{equation}
(compare it to the exact result following from
Eq.~(\ref{H_eff^psi}) shown in Fig.~\ref{Fig8}).


\subsection{\label{III-D}A vortex pinned by the vacancy}


An analogue of the condition (\ref{H_phipsi=0_condition}) for the
field $\psi_{\bf r}$, which would assure that $H^{\varphi,\psi} =
0$ in the diluted Villain model (\ref{H_vill_dil}), reads as
\begin{equation}\label{H_phipsi=0_cond_dil}
c_{\bf r}\sum_{\bf u}\left(\psi_{\bf r} - \psi_{\bf r+u} - 2\pi
m_{\bf r,r+u}\right) c_{\bf r+u}\ =\ 0\ \textrm{ for all }\ {\bf
r}\ .
\end{equation}

\begin{figure}[h]
\includegraphics[width=0.3\textwidth]{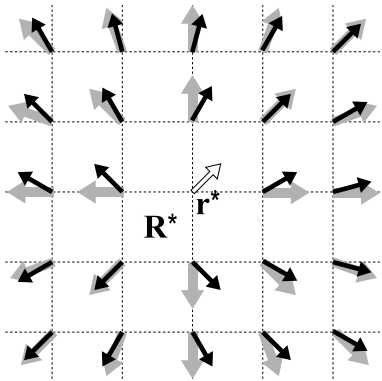}
\caption{Representation field $\psi_{\bf r}$ of topological charge
$q=+1$ situated at site $\bf R^*$ of the dual lattice, which leads
to its decoupling in the Hamiltonian of the pure Villain model
(black and white arrows) and a model with a vacancy at site $\bf
r^*$ (grey arrows).}\label{Fig9}
\end{figure}

Numerical studies of spin vortices in the presence of a spinless
site \cite{Wysin03}, \cite{Pereira03} suggest that it is
energetically preferable for a vortex to be pinned (centered) on
the vacancy. Thus, one can assume that
\begin{equation}\label{vort_field_pinned}
\widetilde{\psi}_{\bf r}\ =\ \pm\Phi_{\bf r}({\bf r}^*)\ ,
\end{equation}
where $\Phi_{\bf r}({\bf r}^*)$ was defined after Eq.
(\ref{vort_field}) and ${\bf r}^*$ is the coordinate of the
vacancy, might satisfy (\ref{H_phipsi=0_cond_dil}) when the
topological charge $q=\pm 1$ is on one of the four dual lattice
sites $\bf R^*$ adjacent to ${\bf r}^*$ (see Fig. \ref{Fig9}).


Then, the vortex-on-vacancy pinning energy, i.e. the energy of the
vortex centered on $\bf r^*$ minus that of the vortex centered on
$\bf R^*$, can be estimated by numerical summation over a lattice
of sufficiently large size:
\begin{eqnarray}\nonumber
E_{\mathrm{pin}} = \frac{J}{2} \sum_{\left<{\bf r,r'}\right>}
\Big[\left( \widetilde{\psi}_{\bf r} - \widetilde{\psi}_{\bf r'} -
2\pi m_{\bf r,r'} \right)^2
\\\label{E_pin_num}
- \left( \psi'_{\bf r} - \psi'_{\bf r'} - 2\pi m_{\bf r,r'}
\right)^2 \Big] ,
\end{eqnarray}
where $\psi'_{\bf r} = \pm \Phi_{\bf r}({\bf R}^*)$, which gives
$E_{\mathrm{pin}} \simeq - 5.22 J$.

The corresponding energy that follows from (\ref{H_eff^psi}) is
$E_{\mathrm{pin}} = - (3\pi - 4) J \simeq - 5.42 J$. The
difference from the result of (\ref{E_pin_num}) is not surprising,
if one notices that $\widetilde{\psi}_{\bf r}$ only approximately
fulfills (\ref{H_phipsi=0_cond_dil}) for almost all the lattice
sites.

It is worth mentioning that using the exchange potential of the
$2D$ $XY$ model, Eq. (\ref{H_XY_0}), numerical summation analogous
to that of (\ref{E_pin_num}) leads to $E^{2DXY}_{\mathrm{pin}}
\simeq - 3.21 J$, which agrees with $-3.178 J$ of the energy
minimizing iterative method \cite{Wysin03} and $-3.54 J$ of the
spin dynamics simulations \cite{Pereira03}.

\section{\label{conclus}Conclusions}

The exact and asymptotic expressions for the interaction energy of
topological charges and a spinless site, Eqs. (\ref{H_eff^psi})
and (\ref{V_1^psi_asympt}), found for the Villain model on a
square lattice, definitively confirm the attractive character of
the interaction. This agrees with the results of the spin dynamics
simulations for the $2D$ $XY$ model \cite{Pereira03} and the
energy minimizing iterative method for the easy-plane Heisenberg
model \cite{Wysin03}.

However, we showed that this interaction in the $2D$ $XY$ model
can differ from Eq. (\ref{H_eff^psi}), that corresponds to the
Villain model, considerably at small separations due to different
energies of the vortex ``cores'' (regions with strong
disorientation of spins). In particular, the exact value of the
vortex-on-vacancy pinning energy in the Villain model,
$E_{\mathrm{pin}} = - (3\pi - 4) J \simeq - 5.42 J$, differs
significantly from that found in \cite{Pereira03} and
\cite{Wysin03} ($-3.54 J$ and $-3.178 J$, respectively).

Moreover, we showed that the mentioned difference of the vortex
cores' energies in the two models leads to a deviation of the
vortex-antivortex interaction energy in the $2D$ $XY$ model from a
logarithmic law at small separations, while the corresponding
energy of the Villain model retains logarithmic dependence on
separation $x$ (if we neglect slight anisotropy effects) up to the
smallest possible distance on a lattice which is of one lattice
spacing, $x=a$.

We have estimated the vortex-antivortex pair creation energy for
the $2D$ $XY$ model as $E^{2DXY}_{\mathrm{pair}}(a) \simeq 6.6 J$
(in contrast to $\pi^2 J \simeq 9.9 J$ of the Villain model),
which is in reasonable agreement with the results of the recent
Monte Carlo simulations \cite{GuptaBaillie92}, \cite{OtaOta95}
($7.55 J$ and $7.3 J$, respectively).

\begin{acknowledgments}
We thank Prof. Bertrand Berche for his useful corrections to the
manuscript, and for the discussions we have had. We wish to
acknowledge the support of cooperation programme ``Dnipro''
between the Ministry of Foreign Affairs of France and the Ministry
of Education and Science of Ukraine. Yu. H. was supported in part
by the Austrian Fonds zur F\"orderung der wissenschaftlichen
Forschung under Project No. P19583-N20.
\end{acknowledgments}

\appendix

\section{\label{appendxA}Functions $I_{sc}(X,Y)$ and $I_{ss}(X,Y)$}

In this appendix we study the functions
\begin{equation}\label{Isc_def}
I_{sc}(X,Y) = \frac{1}{N} \sum_{\bf k}
\frac{\sin\frac{k_xa}{2}\cos\frac{k_ya}{2}}{\sum_{\alpha=x,y}\sin^2\frac{k_\alpha
a}{2}} \sin k_x X \cos k_y Y\ ,
\end{equation}
\begin{equation}\label{Iss_def}
I_{ss}(X,Y) = \frac{1}{N} \sum_{\bf k}
\frac{\sin\frac{k_xa}{2}\sin\frac{k_ya}{2}}{\sum_{\alpha=x,y}\sin^2\frac{k_\alpha
a}{2}} \sin k_x X \sin k_y Y\ ,
\end{equation}
which enter many important expressions concerning the behavior of
topological charges, and the sums over $\bf k$ in (\ref{Isc_def}),
(\ref{Iss_def}) span the first Brillouin zone.

For $X=(2n-1)a$, $Y=(2m-1)a$ ($n,m = 1,2,3,\ldots$)
(\ref{Isc_def}) and (\ref{Iss_def}) can be calculated exactly,
replacing the sums with integrals in the thermodynamic limit. The
integration gives
\begin{widetext}
\begin{eqnarray}\nonumber
I_{sc}((2n-1)a,(2m-1)a) &=& \sum_{i=0}^{n-1}
\frac{(-1)^i(2n-1)!}{(2(n-i-1))!(2i+1)!} \sum_{j=0}^{m-1}
\frac{(-1)^j(2m-1)!}{(2(m-j)-1)!(2j)!}\hspace{-0.1cm}
\sum_{k=0}^{n-i-1} \frac{(-1)^k(n-i-1)!}{(n-i-k-1)! k!}
\\\label{Isc_exact} && \times \sum_{l=0}^{m-j}
\frac{(-1)^l(m-j)!}{(m-j-l)!\ l!}\ F(i+k+1,j+l)\ ,
\end{eqnarray}
\begin{eqnarray}\nonumber
I_{ss}((2n-1)a,(2m-1)a) &=& \sum_{i=0}^{n-1}
\frac{(-1)^i(2n-1)!}{(2(n-i-1))!(2i+1)!} \sum_{j=0}^{m-1}
\frac{(-1)^j(2m-1)!}{(2(m-j-1))!(2j+1)!} \sum_{k=0}^{n-i-1}
\frac{(-1)^k(n-i-1)!}{(n-i-k-1)!\ k!}
\\\label{Iss_exact} && \times \sum_{l=0}^{m-j-1} \frac{(-1)^l(m-j-1)!}{(m-j-l-1)!\ l!}
\ F(i+k+1,j+l+1)
\end{eqnarray}
with
\begin{eqnarray}\nonumber
F(p,q) &=& \sum_{u=0}^{q-1} (-1)^u
\frac{(2(p+u)-1)!!}{(2(p+u))!!}\
\frac{(2(q-u-1)-1)!!}{(2(q-u-1))!!}\ +\
\frac{1}{2}\sum_{u=0}^{p+q-1}
\frac{(-1)^{q+u}(p+q-1)!(2u-1)!!}{(p+q-u-1)!\ (u!)^2}
\\\label{F(p,q)} &&-\ \frac{1}{\pi} \sum_{u=1}^{p+q-1}
\frac{(-1)^{q+u}(p+q-1)}{(p+q-u-1)!\ u!} \left(
\frac{(2u-1)!!}{u!} \sum_{w=1}^{u-1}\frac{(u-w-1)!}{(2(u-w)-1)!!}
+ \frac{1}{u} \right).
\end{eqnarray}
\end{widetext}

These results were obtained by expressing $\sin k_x X$, $\sin k_y
Y$, $\cos k_y Y$ as polynomials $P\left(\sin \frac{k_\alpha a}{2},
\cos \frac{k_\alpha a}{2}\right)$, and then applying the standard
tables of integrals~\cite{Prudnikov}. We used the notations:
$(2n)!!\equiv \prod_{i=1}^{n} 2i$, $(2n-1)!!\equiv \prod_{i=1}^{n}
(2i-1)$; when $n=0$: $(2n)!!\equiv 1$, $(2n-1)!!\equiv 1$. The
sums of no meaning, like $\sum_{i=n}^{m}$ with $m<n$, that may be
encountered in (\ref{F(p,q)}) for some values of $p$, $q$, should
be interpreted as equal to zero.

It is instructive to find an asymptotic form for (\ref{Isc_def})
and (\ref{Iss_def}). It turns out that simple analytic expressions
can be obtained, assuming that at least one of the arguments $X,Y$
is large. Using the integral \cite{Tallarida}
\begin{equation}
\int_{0}^{\infty} \frac{\cos x}{x^2 + a^2} = \frac{\pi}{2|a|}
e^{-|a|} ,
\end{equation}
one can show that
\begin{eqnarray}\nonumber
I_{sc}(X\to\infty,Y) &=& \frac{a}{\pi} \int_{0}^{\pi/a} dk_y
e^{-X\frac{2}{a}\sin\frac{k_ya}{2}}\cos k_y Y
\\\nonumber
&&\times \sinh\left(\sin\frac{k_ya}{2}\right)\cot \frac{k_ya}{2}
\\\label{Isc_1}
&=& \frac{a}{\pi} \int_{0}^{\pi/a} dk_y e^{-X k_y}\cos k_y Y
\end{eqnarray}
and
\begin{eqnarray}\nonumber
I_{sc}(X,Y\to\infty) = \frac{a}{\pi} \int_{0}^{\pi/a} dk_x
e^{-Y\frac{2}{a}\sin\frac{k_xa}{2}} \cos k_x X
\\\label{Isc_2}
\times \cosh\left(\sin\frac{k_xa}{2}\right) = \frac{a}{\pi}
\int_{0}^{\pi/a} dk_y e^{-Y k_x}\sin k_x X .\quad
\end{eqnarray}
So,
\begin{equation}\label{Isc_asympt}
I_{sc}(X,Y) = \frac{a}{\pi} \frac{X}{X^2 + Y^2},
\end{equation}
when at least one of the arguments $X,Y$ is sufficiently large.

In a similar way one can show that
\begin{equation}\label{Iss_asympt}
I_{ss}(X,Y) = \frac{a^2}{\pi} \frac{XY}{\left(X^2 + Y^2\right)^2},
\end{equation}
if at least one of the arguments $X,Y$ is sufficiently large.

\section{\label{appendxB}Hamiltonian describing topological charges in a system with
         a spin vacancy}

The aim of the present appendix is to show how the ``vortexless''
degrees of freedom $\varphi_{\bf r}$ can be integrated out in the
partition function (\ref{Z_0}) when only one spin vacancy at site
$\bf r^*$ is considered, so that (see Subsection \ref{III-A})
\begin{equation}
c_{\bf r}= \Big\{ \begin{array}{ll} 0,\quad {\bf r} = {\bf r}^*, \\
1,\quad {\bf r} \neq {\bf r}^*;
\end{array} \qquad\textrm{or}\qquad p_{\bf r}= \Big\{ \begin{array}{ll} 1,\quad {\bf r} = {\bf r}^*, \\
0,\quad {\bf r} \neq {\bf r}^*.
\end{array}
\end{equation}

As we will show below, the partition function can be presented in
this case in the form (\ref{Z_1}) with
\begin{eqnarray}\nonumber
&& H_{\mathrm{eff}} = H_{\psi} - \frac{1}{4\beta^2 J}\Bigg(
\sum_{\bf k} \frac{\eta_{\bf k}\eta_{\bf -k}}{\gamma_{\bf k}}
\\\nonumber
&&- \frac{\pi}{4(\pi-2)} \frac{1}{N} \sum_{\bf k,k'}(g_{\bf k,-k'}
+ g_{\bf k,k'})\frac{\eta_{-\bf k}\eta_{-\bf k'}}{\gamma_{\bf
k'}}e^{-i({\bf k+k'}){\bf r}^*}
\\\label{H_eff}
&&+ \frac{\pi}{4}\frac{1}{N} \sum_{\bf k,k'}(g_{\bf k,-k'} +
g_{\bf k,k'})\frac{\eta_{-\bf k}\eta_{-\bf k'}}{\gamma_{\bf
k'}}e^{-i({\bf k+k'}){\bf r}^*} \Bigg) ,
\end{eqnarray}
where $\gamma_{\bf k} \equiv 2\left(\sin^2\frac{k_xa}{2} +
\sin^2\frac{k_ya}{2}\right)$,
\begin{equation}\label{g_def}
g_{\bf k,k'} \equiv (\gamma_{{\bf k} + {\bf k}'} - \gamma_{\bf k}
- \gamma_{{\bf k}'})/\gamma_{\bf k},
\end{equation}
and
\begin{equation}\label{eta_def}
\eta_{\bf k} \equiv \frac{\beta J}{\sqrt{N}}\ e^{-i{\bf k r}^*}
\sum_{\bf u} e^{-i{\bf k u}} (\psi_{\bf r^*+u} - \psi_{\bf r^*} -
2\pi m_{{\bf r^*+u},{\bf r^*}}).
\end{equation}

\subsection{\label{appendxB-1}The partition function of the Villain model on the lattice
            with a spin vacancy}

Let us denote
\begin{equation}
Z_\psi \equiv \mathrm{Tr}_\varphi e^{-\beta (H_\varphi +
H_{\varphi,\psi})},
\end{equation}
so
\begin{equation}\label{Z}
Z = \mathrm{Tr}_\psi \left(e^{-\beta H_\psi} Z_\psi\right).
\end{equation}

Using Fourier transformation (\ref{phi_Fourier}), one can rewrite
the terms that depend on $\varphi_{\bf r}$ in the Hamiltonian
(\ref{H_phi+H_psi+H_phipsi}) as (see \cite{Kapikranian07}):
\begin{equation}
H_\varphi = J \sum_{\bf k} \gamma_{\bf k} \varphi_{\bf k}
\varphi_{\bf -k} + \frac{J}{N}\sum_{\bf k,k'}e^{-i({\bf k+k'}){\bf
r^*}} g_{\bf k,k'} \varphi_{\bf k} \varphi_{\bf k'},\qquad
\end{equation}
where $g_{\bf k,k'}$ was defined in (\ref{g_def}) and the sums are
over the 1st Brillouin zone. Correspondingly, the mixed
$\varphi\psi$-term in Eq.~(\ref{H_phi+H_psi+H_phipsi}) reads:
\begin{eqnarray}\nonumber
H_{\varphi,\psi} &=& - \frac{J}{\sqrt{N}} \sum_{\bf k}
\Bigg(\varphi_{\bf k} e^{-i{\bf k r}^*}
\\\label{H_phipsi}
&&\times\sum_{\bf u} e^{-i{\bf k u}} (\psi_{\bf r^*+u} - \psi_{\bf
r^*} - 2\pi m_{{\bf r^*+u},{\bf r^*}}) \Bigg).\qquad
\end{eqnarray}

Using the Taylor series expansion, $Z_\psi$ can be written as:
\begin{eqnarray}\nonumber
Z_\psi &=& \mathrm{Tr}_\varphi e^{-\beta J \sum_{\bf k}
\gamma_{\bf k} \varphi_{\bf k} \varphi_{\bf -k} + \sum_{\bf k}
\eta_{\bf k} \varphi_{\bf k}} \\\label{Z_psi_1} && \times \left(1
+ \sum_{n=1}^{\infty} \frac{1}{n!} I_{(\varphi_{{\bf
k}_1},\varphi_{{\bf k}_2}),\ldots (\varphi_{{\bf
k}_{2n-1}},\varphi_{{\bf k}_{2n}})} \right)\qquad
\end{eqnarray}
where $\eta_{\bf k}$ was defined in (\ref{eta_def}) and
\begin{eqnarray}\nonumber
I_{(\varphi_{{\bf k}_1},\varphi_{{\bf k}_2}),\ldots (\varphi_{{\bf
k}_{2n-1}},\varphi_{{\bf k}_{2n}})} \equiv \frac{(-\beta
J)^n}{N^n} \sum_{{\bf k}_1,{\bf k}_2}\cdots \sum_{{\bf
k}_{2n-1},{\bf k}_{2n}}
\\\label{Idef} \times\ e^{-i({\bf k}_1 + \ldots + {\bf k}_{2n}){\bf
r}^*} g_{{\bf k}_1, {\bf k}_2} \cdots g_{{\bf k}_{2n-1}, {\bf
k}_{2n}} \varphi_{{\bf k}_1} \cdots \varphi_{{\bf k}_{2n}} .\quad\
\end{eqnarray}

Now, introducing the notations:
\begin{equation}
Z_*\ \equiv\ \mathrm{Tr}_\varphi e^{-\beta J \sum_{\bf k}
\gamma_{\bf k} \varphi_{\bf k} \varphi_{\bf -k} + \sum_{\bf k}
\eta_{\bf k} \varphi_{\bf k}}
\end{equation}
and
\begin{equation}\label{avrg_*}
\left<\ldots\right>_*\ \equiv\ Z_*^{-1}\ \mathrm{Tr}_\varphi
\left( e^{-\beta J \sum_{\bf k} \gamma_{\bf k} \varphi_{\bf k}
\varphi_{\bf -k} + \sum_{\bf k} \eta_{\bf k} \varphi_{\bf
k}}\ldots\right)\ ,
\end{equation}
(\ref{Z_psi_1}) becomes
\begin{equation}\label{Z_psi_2}
Z_\psi = Z_* \left(1 + \sum_{n=1}^{\infty} \frac{1}{n!}
\left<I_{(\varphi_{{\bf k}_1},\varphi_{{\bf k}_2}),\ldots
(\varphi_{{\bf k}_{2n-1}},\varphi_{{\bf k}_{2n}})}\right>_*
\right) .
\end{equation}

\subsection{\label{appendxB-2}Calculation of $Z_*$ and $\left<\varphi_{{\bf k}_1}
            \ldots \varphi_{{\bf k}_{2n}}\right>_*$}

$Z_*$ and $\left<\varphi_{{\bf k}_1} \ldots \varphi_{{\bf
k}_{2n}}\right>_*$ are the first quantities to be calculated.

Since $\varphi_{\bf k}$ (for ${\bf k} \neq 0$) is a complex
quantity: $\varphi_{\bf k} = \varphi^c_{\bf k} + i \varphi^s_{\bf
k}$, $\mathrm{Tr}_{\varphi}$ should be understood as:
\begin{equation}\label{Tr_phi}
\mathrm{Tr}_{\varphi} = \prod_{{\bf k}\in
B_{1/2}}\int_{-\infty}^\infty d\varphi^c_{\bf k}
\int_{-\infty}^\infty d\varphi^s_{\bf k}\ ,
\end{equation}
where $B_{1/2}$ stands for a half of the 1st Brillouin zone
excluding ${\bf k} = 0$ ($\varphi^c_{\bf k}$ and $\varphi^s_{\bf
k}$ in the other half are not independent, due to the relations:
$\varphi^c_{\bf -k} = \varphi^c_{\bf k}$ and $\varphi^s_{\bf -k} =
-\varphi^s_{\bf k}$). It was possible to extend the bounds of
integration to infinity in (\ref{Tr_phi}) and omit writing the
integral over $\varphi_0$, since the functions that stand after
the trace in our calculations are always rapidly decaying when
$\beta J\to\infty$ and independent from $\varphi_0$.

Then, it is straightforward to obtain
\begin{equation}\label{Z_*_1}
Z_*\ =\ \left(\prod_{{\bf k} \neq 0}\sqrt{\frac{\pi}{2\beta
J\gamma_{\bf r}}}\right) e^{\frac{1}{4\beta J}\sum_{{\bf k}\neq
0}\frac{\eta_{\bf k} \eta_{\bf -k}}{\gamma_{\bf k}}}
\end{equation}
Using (\ref{Z_*_1}), it is easy to show that
\begin{equation}
\left<\varphi_{{\bf k}_1} \ldots \varphi_{{\bf k}_{2n}}\right>_* =
Z_*^{-1} 2^{-2n} \frac{\partial}{\partial\eta_{{\bf k}_1}} \cdots
\frac{\partial}{\partial\eta_{{\bf k}_{2n}}}\ Z_*\ ,
\end{equation}
where
\begin{equation}
\frac{\partial}{\partial\eta_{\bf k}} \equiv
\frac{\partial}{\partial\eta^c_{\bf k}} - i
\frac{\partial}{\partial\eta^s_{\bf k}}\ ,\quad
\frac{\partial}{\partial\eta_{\bf -k}} \equiv
\frac{\partial}{\partial\eta^c_{\bf k}} + i
\frac{\partial}{\partial\eta^s_{\bf k}}\ .
\end{equation}
Noting that $\frac{\partial\eta_{\bf k}}{\partial\eta_{\bf k'}} =
2 \delta_{\bf k,k'}$ ($\delta_{\bf k,k'}$ is Kronecker delta), one
arrives at
\begin{eqnarray}\nonumber
\left<\varphi_{{\bf k}_1} \ldots \varphi_{{\bf k}_{2n}}\right>_*
&=& \sum_{l=0}^{n} \frac{1}{(2\beta J)^{2n-l}}
\sum_{\mathrm{pairs}\ 2n\to l} \\\label{phi...phi_avrg_*} &&
\times \prod_{u=1}^{l} \frac{\delta_{{\bf k}_{i_u},-{\bf
k}_{j_u}}}{\gamma_{{\bf k}_{i_u}}} \prod_{w=1}^{2n-2l}
\frac{\eta_{-{\bf k}_{p_w}}}{\gamma_{{\bf k}_{p_w}}}\ ,\qquad
\end{eqnarray}
where the sum $\displaystyle \sum_{\mathrm{pairs}\ 2n\to l}$ spans
all the possible ways of selecting $l$ indistinguishable unordered
pairs $(i_u,j_u),\ u=1,\ldots ,l$ out of $2n$ indexes $1,\ldots ,
2n$. (It is easy to see that
\begin{equation}\nonumber
\sum_{\mathrm{pairs}\ 2n\to l} 1\ =\
\frac{(2n)!}{l!(2!)^l(2n-2l)!}\ .\ \Bigg)
\end{equation}

\subsection{\label{appendxB-3}Calculation of $\left<I_{(\varphi_{{\bf k}_1},\varphi_{{\bf k}_2}),\ldots
(\varphi_{{\bf k}_{2n-1}},\varphi_{{\bf k}_{2n}})}\right>_*$}

According to (\ref{Idef}) and (\ref{avrg_*}),
\begin{eqnarray}\nonumber
\left< I_{(\varphi_{{\bf k}_1},\varphi_{{\bf k}_2}),\ldots
(\varphi_{{\bf k}_{2n-1}},\varphi_{{\bf k}_{2n}})} \right>_* =
\frac{(-\beta J)^n}{N^n} \sum_{{\bf k}_1, \ldots {\bf k}_{2n}}
\\\nonumber \times e^{-i({\bf k}_1 + \ldots + {\bf k}_{2n}){\bf
r}^*} g_{{\bf k}_1, {\bf k}_2} \cdots g_{{\bf k}_{2n-1}, {\bf
k}_{2n}}\left< \varphi_{{\bf k}_1} \cdots \varphi_{{\bf k}_{2n}}
\right>_* .\\\label{I_avrg_*}
\end{eqnarray}

At this stage, it is convenient to introduce the notions:
\begin{equation}\label{I_i_def}
I_i \equiv \frac{1}{N^i} \sum_{{\bf k}_1,\ldots, {\bf k}_i}
g_{{\bf k}_1, -{\bf k}_2} g_{{\bf k}_2, -{\bf k}_3} \cdots g_{{\bf
k}_{i-1}, -{\bf k}_i} g_{{\bf k}_i, -{\bf k}_1}\ ,
\end{equation}
\begin{eqnarray}\nonumber
I^*_i & \equiv& \frac{1}{N^i} \sum_{{\bf k}_1,\ldots, {\bf
k}_{i+1}} g_{{\bf k}_1, -{\bf k}_2} g_{{\bf k}_2, -{\bf k}_3}
\cdots g_{{\bf k}_{i-1}, -{\bf k}_i} g_{{\bf k}_i, {\bf
k}_{i+1}}\\\label{I*_i_def} &&\times\ e^{-i({\bf k}_1 + {\bf
k}_{i+1}){\bf r^*}}\ \frac{\eta_{{\bf k}_1} \eta_{{\bf
k}_{i+1}}}{\gamma_{{\bf k}_{i+1}}}\ .
\end{eqnarray}

Then, insertion of (\ref{phi...phi_avrg_*}) into (\ref{I_avrg_*})
leads to a polynomial form with respect to $I_i$ and $I^*_i$
($i=1,\ldots,\infty$):
\begin{eqnarray}\nonumber
&& \left< I_{(\varphi_{{\bf k}_1},\varphi_{{\bf k}_2}),\ldots
(\varphi_{{\bf k}_{2n-1}},\varphi_{{\bf k}_{2n}})} \right>_* =
(-1)^n \sum_{l=0}^{n} \frac{2^{-n}}{(2\beta J)^{n-l}} \\\nonumber
&& \times \left(\prod_{i=1}^{l}\sum_{\lambda_i = 0}^{[l/i]}\right)
\left(\prod_{j=1}^{n-l}\sum_{\lambda^*_j = 0}^{[(n-l)/j]}\right)
\delta\left(\sum_{i=1}^{l}i\lambda_i -l\right) \\\nonumber &&
\times\ \delta\left(\sum_{j=1}^{n-l}j\lambda^*_j -(n-l)\right)
\Lambda_{\lambda_1,\ldots, \lambda_{l}}^{\lambda^*_1,\ldots,
\lambda^*_{n-l}} I_1^{\lambda_1}\cdots I_l^{\lambda_l}
\\\label{I_avrg_*_1} && \times\ (I^*_1)^{\lambda^*_1}\cdots
(I^*_{n-l})^{\lambda^*_{n-l}}\ ,
\end{eqnarray}
where $[a]$ means the nearest integer not exceeding $a$,
$\delta(x) = \Big\{ \begin{array}{ll} 1,\ x=0 \\ 0,\ x\neq 0
\end{array}$, and
\begin{equation}\label{Lambda}
\Lambda_{\lambda_1,\ldots, \lambda_{l}}^{\lambda^*_1,\ldots,
\lambda^*_{n-l}} = n! \prod_{i=1}^{l} \frac{\left[2^{i-1}
(i-1)!\right]^{\lambda_i}}{\lambda_i! (i!)^{\lambda_i}}
\prod_{j=1}^{n-l} \frac{\left[2^{j-1}
j!\right]^{\lambda^*_j}}{\lambda^*_i! (i!)^{\lambda^*_i}}
\end{equation}
is the combinatorial ``weight'' given by the number of ways of
selecting $\lambda_1$ unordered elements, $\lambda_2$ unordered
groups of two unordered elements, ..., $\lambda_l$ unordered
groups of $l$ unordered elements, $\lambda^*_1$ unordered
elements, $\lambda^*_2$ unordered groups of two unordered
elements, ..., and $\lambda^*_{n-l}$  unordered groups of $n-l$
unordered elements out of $n$ distinct elements, which is
\begin{eqnarray}\nonumber
n!/(\lambda_1! \lambda_2!\cdots\lambda_l!\ \lambda^*_1!
\lambda^*_2!\cdots\lambda^*_{n-l}!\ (1!)^{\lambda_1}
(2!)^{\lambda_2} \cdots (l!)^{\lambda_l}
\\\nonumber
(1!)^{\lambda^*_1}(2!)^{\lambda^*_2} \cdots
((n-l)!)^{\lambda^*_{n-l}})\ ,
\end{eqnarray}
times the number of distinct ways of connecting 4 distinct
elements belonging to 2 distinct groups, each consisting of two
elements, with 2 indistinguishable links in such a manner that the
two elements of one group are connected to the elements belonging
to another group, raised to the power ${\lambda_2}$, times the
product over $i=\overline{1,l}$ of the number of distinct ways of
connecting $2i$ distinct elements belonging to $i$ distinct
groups, each consisting of two elements, with $i$
indistinguishable links in such a manner that the two elements of
each group are connected to the elements belonging to two another
groups, raised to the power ${\lambda_i}$, i.e.
\begin{equation}
\times \prod_{i=1}^{l} \left[2^{i-1} (i-1)!\right]^{\lambda_i}\ ,
\end{equation}
times the number of distinct ways of connecting 4 distinct
elements belonging to 2 distinct groups, each consisting of two
elements, with 1 link in such a manner that one of the two
elements of one group is connected to one of the two elements
belonging to another group, raised to the power $\lambda_2^*$,
times the product over $j=\overline{1,n-l}$ of the number of
distinct ways of connecting $2j$ distinct elements belonging to
$j$ distinct groups, each consisting of two elements, with $j-1$
indistinguishable links in such a manner that one of the two
elements of any group is connected to one of the two elements of
another group and the second element is either connected to one of
the two elements of a different group or not connected, raised to
the power $\lambda^*_{j}$, i.e.
\begin{equation}
\times \prod_{j=1}^{n-l} \left[2^{j-1} j!\right]^{\lambda^*_j} .
\end{equation}

Inserting (\ref{Lambda}) into (\ref{I_avrg_*_1}), one has
\begin{eqnarray}\nonumber
&& \left< I_{(\varphi_{{\bf k}_1},\varphi_{{\bf k}_2}),\ldots
(\varphi_{{\bf k}_{2n-1}},\varphi_{{\bf k}_{2n}})} \right>_* =
(-1)^n n! \sum_{l=0}^{n} \frac{1}{(2\beta J)^{n-l}}
\\\nonumber && \times \prod_{i=1}^{l}\sum_{\lambda_i =
0}^{[l/i]} \frac{1}{\lambda_i!}\left( \frac{I_i}{2i}
\right)^{\lambda_i} \prod_{j=1}^{n-l}\sum_{\lambda^*_j =
0}^{[(n-l)/j]} \frac{1}{\lambda^*_i!}\left( \frac{I^*_i}{2}
\right)^{\lambda^*_i}
\\\label{I_avrg_*_2} &&
\times\ \delta\left(\sum_{i=1}^{l}i\lambda_i -l\right)
\delta\left(\sum_{j=1}^{n-l}j\lambda^*_j -(n-l)\right)\ ,
\end{eqnarray}
and then, inserting (\ref{I_avrg_*_2}) in (\ref{Z_psi_2}), one can
notice that the infinite series in (\ref{Z_psi_2}) can be
rearranged as it is shown below:
\begin{eqnarray}\nonumber
Z_\psi &=& Z_* \prod_{i=1}^{\infty}\Bigg(1 + (-1)^i\frac{I_i}{2i}
+ \frac{1}{2!} \left((-1)^i\frac{I_i}{2i}\right)^2 \\\nonumber &&
+ \frac{1}{3!} \left((-1)^i\frac{I_i}{2i}\right)^3 + \ldots \Bigg)
\prod_{j=1}^{\infty}\Bigg(1 + (-1)^j\frac{I_j^*}{4\beta J}
\\\nonumber && + \frac{1}{2!} \left((-1)^j\frac{I^*_j}{4\beta
J}\right)^2+ \frac{1}{3!} \left((-1)^j\frac{I^*_j}{4\beta
J}\right)^3 + \ldots \Bigg)\ ,
\end{eqnarray}
or
\begin{eqnarray}\nonumber
Z_\psi &=& Z_* \exp\left({\frac{1}{2}\sum_{i=1}^{\infty}(-1)^i
I_i/i}\right) \\\label{Z_psi_3} &&\times
\exp\left({\frac{1}{4\beta J}\sum_{j=1}^{\infty}(-1)^j
I^*_j}\right) .
\end{eqnarray}

\subsection{\label{appendxB-4}Calculation of $I_i$ and $I^*_i$}

Eqs.~(\ref{I_i_def}) and (\ref{I*_i_def}) can be written as
\begin{equation}
I_i = \frac{1}{N} \sum_{\bf k} \tilde{I}_{i-1} ({\bf k,-k})
\end{equation}
and
\begin{equation}
I^*_i = \frac{1}{N} \sum_{\bf k,k'} \tilde{I}_{i-1} ({\bf k,k}')
\frac{\eta_{-\bf k} \eta_{-\bf k'}}{\gamma_{\bf k'}}\ e^{-i({\bf k
+ k'}){\bf r}^*}
\end{equation}
($i \ge 1$) with
\begin{equation}
\tilde{I}_{i} ({\bf k,k}') \equiv \frac{1}{N^{i}} \sum_{{\bf
k}_1,\ldots, {\bf k}_{i}} g_{{\bf k}, -{\bf k}_1} g_{{\bf k}_1,
-{\bf k}_2} \cdots g_{{\bf k}_{i-1}, -{\bf k}_{i}} g_{{\bf k}_{i},
{\bf k}'}
\end{equation}
for $i \ge 1$ and $\tilde{I}_0({\bf k,k'}) \equiv g_{\bf k,k'}$.
One can notice the obvious recurrent relation
\begin{equation}
\tilde{I}_{i+1} ({\bf k,k}') = \frac{1}{N} \sum_{{\bf k}^*}
\tilde{I}_{i} ({\bf k,-k}^*) g_{{\bf k}^*, {\bf k}'}\ .
\end{equation}

In the thermodynamic limit, one can replace the sum
$\frac{1}{N}\sum_{\bf k}$ over the 1st Brillouin zone by the
integrals $\frac{a^2}{(2\pi)^2} \int_{-\pi/a}^{\pi/a} dk_x
\int_{-\pi/a}^{\pi/a} dk_y$, and then, noticing that
\begin{equation}\nonumber
\frac{a^2}{\pi^2}\int_{0}^{\pi/a} dk_x \int_{0}^{\pi/a} dk_y
\frac{\sin^4\frac{k_xa}{2}}{\sin^2\frac{k_xa}{2} +
\sin^2\frac{k_ya}{2}}\ =\ \frac{1}{\pi}
\end{equation}
and
\begin{eqnarray}\nonumber
\frac{a^2}{\pi^2}\int_{0}^{\pi/a} dk_x \int_{0}^{\pi/a} dk_y
\frac{\sin^2\frac{k_xa}{2}\cos^2\frac{k_xa}{2}}{\sin^2\frac{k_xa}{2}
+ \sin^2\frac{k_ya}{2}} = \frac{a^2}{\pi^2} \\\nonumber \times
\int_{0}^{\pi/a} dk_x \int_{0}^{\pi/a} dk_y
\frac{\sin^2\frac{k_xa}{2}\sin^2\frac{k_ya}{2}}{\sin^2\frac{k_xa}{2}
+ \sin^2\frac{k_ya}{2}} =\ \frac{1}{2} - \frac{1}{\pi}\ ,
\end{eqnarray}
one can show that
\begin{eqnarray}\nonumber
\frac{1}{N}\sum_{\bf k'} g_{\bf k,-k'} g_{\bf k',k''} =
\left(1-\frac{2}{\pi}\right) g_{\bf k,-k''}
\\\nonumber
- \frac{1}{\pi} \left(g_{\bf k,-k''} + g_{\bf k,k''}\right) +
\left(\frac{1}{2}-\frac{1}{\pi}\right) \gamma_{\bf k''}\ ,
\end{eqnarray}
\begin{eqnarray}\nonumber
\frac{1}{N}\sum_{\bf k'} g_{\bf k,k'} g_{\bf k',k''} =
\left(1-\frac{2}{\pi}\right) g_{\bf k,k''}
\\\nonumber
- \frac{1}{\pi} \left(g_{\bf k,-k''} + g_{\bf k,k''}\right) +
\left(\frac{1}{2}-\frac{1}{\pi}\right) \gamma_{\bf k''}\ .
\end{eqnarray}
and
\begin{equation}\nonumber
\frac{1}{N}\sum_{\bf k} \gamma_{\bf k} g_{\bf k,k'} = -
\gamma_{\bf k'}\ .
\end{equation}
Then, it is easy to see that:
\begin{eqnarray}\nonumber
\tilde{I}_{i} ({\bf k,k}') = A_i g_{{\bf k},(-1)^i{\bf k}'} + B_i
\left(g_{\bf k,-k'} + g_{\bf k,k'}\right) + C_i \gamma_{\bf k'}
\end{eqnarray}
with coefficients $A_i$, $B_i$ and $C_i$, obeying the recurrent
relations:
\begin{equation}\nonumber
A_{i+1} = \left(1-\frac{2}{\pi}\right) A_i\ ,
\end{equation}
\begin{equation}\nonumber
B_{i+1} = -\frac{1}{\pi} A_i + \left(1-\frac{4}{\pi}\right) B_i\ ,
\end{equation}
\begin{equation}\nonumber
C_{i+1} = \left(\frac{1}{2}-\frac{1}{\pi}\right) \left(A_i + 2
B_i\right) - C_i\ ,
\end{equation}
and $A_0 = 1$, $B_0 = 0$, $C_0 = 0$. Thus,
\begin{equation}\nonumber
A_{i} = \left(1-\frac{2}{\pi}\right)^i\ ,
\end{equation}
\begin{eqnarray}\nonumber
B_{i} &=& -\frac{1}{\pi} \sum_{j=0}^{i-1}
\left(1-\frac{4}{\pi}\right)^{j}
\left(1-\frac{2}{\pi}\right)^{i-1-j} \\\nonumber &=&
-\frac{1}{2}\left[\left(1-\frac{2}{\pi}\right)^{i} -
\left(1-\frac{4}{\pi}\right)^{i}\right]\ ,
\end{eqnarray}
\begin{eqnarray}\nonumber
C_{i} &=& (-1)^{i-1}\left(\frac{1}{2}-\frac{1}{\pi}\right)
\sum_{j=0}^{i-1} (-1)^{j} \left(1-\frac{4}{\pi}\right)^{j}
\\\nonumber
&=& \frac{1}{4} \left[ (-1)^{i-1} +
\left(1-\frac{4}{\pi}\right)^{i} \right]\ .
\end{eqnarray}

Finally, one can obtain expressions for $I_i$ and $I^*_i$ and
check that
\begin{eqnarray}\nonumber
&& \sum_{i=1}^{\infty} (-1)^i I^*_i = - \frac{\pi}{4(\pi-2)}
\\\nonumber
&& \times \frac{1}{N} \sum_{\bf k,k'}(g_{\bf k,-k'} + g_{\bf
k,k'})\frac{\eta_{-\bf k}\eta_{-\bf k'}}{\gamma_{\bf
k'}}e^{-i({\bf k+k'}){\bf r}^*}
\\\label{sum_I*_i}
&&+\ \frac{\pi}{4}\frac{1}{N} \sum_{\bf k,k'}(g_{\bf k,-k'} +
g_{\bf k,k'})\frac{\eta_{-\bf k}\eta_{-\bf k'}}{\gamma_{\bf
k'}}e^{-i({\bf k+k'}){\bf r}^*}.\qquad
\end{eqnarray}

In conclusion, using (\ref{Z}), (\ref{Z_*_1}), (\ref{Z_psi_3}),
and (\ref{sum_I*_i}), we obtain (\ref{H_eff}).


\end{document}